\documentclass[24pt]{article}
\usepackage{amssymb}
\usepackage{amsfonts}
\usepackage{amssymb,amsmath}
\usepackage{mathrsfs}
\usepackage{latexsym}
\usepackage{amsmath}
\usepackage{latexsym}
\usepackage[cp1251]{inputenc}
\usepackage{graphicx}
\usepackage{color}

\date{}

\title{Reformulation of a transmission and reflection problems in terms of a quantum wave impedance function }
\author{O. I. Hryhorchak\\
{\small Department for Theoretical Physics, Ivan Franko National
University of Lviv,}\\
{\small 12, Drahomanov Str., Lviv, UA--79005,
Ukraine}\\
\small{\it{Orest.Hryhorchak@lnu.edu.ua}}}

\def\arcth{\mathop{\rm arcth}\nolimits}
\def\th{\mathop{\rm th}\nolimits}

\def\sh{\mathop{\rm sh}\nolimits}
\def\ch{\mathop{\rm ch}\nolimits}

\begin{document}
\renewcommand{\abstractname}{Abstract}
\maketitle

\begin{abstract}
On the base of a 1D Shr\"{o}dinger equation the non-linear first-order differential equation (Ricatti type) for a quantum wave impedance function was derived. The advantages of this approach were discussed and demonstrated for a case of a single rectangular barrier. Both the scattering and the bound states problem were reformulated in terms of a quantum wave impedance and  its application for solving both these problems was considered. The expressions for a reflection and a transmission coefficient were found on the base of a quantum wave impedance approach. 
\end{abstract}

\section{Introduction}

The term impedance was introduced by Oliver Heaviside \cite{Heaviside:1886} in 1886. In a few years later
Arthur Kennelly represented it with complex \mbox{numbers \cite{Kennely:1893}.} Usually we understand an impedance as some extension of the resistant concept which is applicable in DC circuits to the AC electrical circuits. It has both a real part and an imaginary part. An imaginary part is caused by both the effects of the induction of voltages in conductors by the magnetic fields and the electrostatic storage of charge induced by voltages between conductors. At the same time the resistance forms the real part of the impedance.
	It would be a mistake to state that impedance is only related with the electrical circuits. Impedance is a generalized concept which is widely used in different areas of science from classical mechanics, electricity and acoustics to geology and medicine. Generally, a wave impedance characterizes the resistance of the media to the propagation of wave processes in it. 
For example, the mechanical impedance relates velocities with forces which act on a mechanical system and indicates how much a mechanical system resists motion caused by a harmonic force \cite{Sabanovic_Ohnishi:2011}.
An acoustic impedance is a measure of the resistance which a medium performs to the acoustic flow. This resistance is a reaction on an acoustic pressure applied to a medium \cite{Kinsler_etall:2000}.
Wave impedance of an electromagnetic wave is the resistance which an electromagnetic wave experiences when propagates along transmission line or through a medium including vacuum. Wave impedance can be defined through the ratio between transverse components of the electric and magnetic fields \cite{Guran_Mittra_Moser:1996}.

In all mentioned cases the equations which describe the propagation of wave processes in different media have the same form and it gives us the grounds for the impedance concept introduction in all these cases. And now the question arise: what about a wave function and a Sr\"{o}dinger equation which describes its behavior. Can we talk about the applicability of the wave impedance concept in this case? The answer is definitely yes, because the Sr\"{o}dinger equation has the same form as equations which describe wave processes in mechanics, acoustics, electrical circuits etc. It also means that we can take the results obtained for a wave impedance in other areas of physics and apply them to  quantum mechanical systems. Khondker, Khan and Anwar were the first who did it \cite{Khondker_Khan_Anwar:1988}. They used an analogy with an electrical transmission line to introduce the concept of a quantum wave impedance. 
Further papers demonstrated the efficacy of a quantum wave impedance approach for an analysis of quantum-mechanical structures with a potential which has a complicated spatial structure.

Although this approach is quite good and allows transiting the relations obtained within a transmission line theory to quantum mechanical systems it is still worth to get the relations for a quantum mechanical impedance from the first principles because it will open the other dimensions of a quantum wave impedance method.
 But nor in the paper \cite{Khondker_Khan_Anwar:1988} nor in the further papers (see for example \cite{Anwar_Khondker_Khan:1989, Khondker:1990, Hague_Hague_Khan:1990, Morrisey_Alamb_Khondker:1992, Khondker_Alam:1991, Khondker_Alam:1992, Kaji_Hayata_Koshiba:1992, Nelin:2004, Nelin:2005, Nelin:2006, Nelin:2007_1, Nelin:2009_2, Khatyan_Gindikina_Nelin:2015, Nazarko_etall:2009, Ashby:2016, Nelin_Zinher_Popsui:2017,  Nelin_Shulha_Zinher:2018}) dedicated to a quantum wave impedance, the systematic introduction of this concept and consistent theory based on the first principles were not proposed. The aim of this article is to fill this gap by constructing such a theory starting from the Sr\"{o}dinger equation; to demonstrate an application of a quantum wave impedance for solving both scattering and bound states problems.

\section{A non-linear differential equation for \\a qua\-ntum wave impedance}
We start from the Shr\"odinger equation and rewrite it   in the following way
\begin{eqnarray}
\left(\frac{\psi'(x)}{\psi(x)}\right)'+\left(\frac{\psi'(x)}{\psi(x)}\right)^2=\frac{2m\left(U(x)-E\right)}{\hbar^2}.
\end{eqnarray}
Multiplying both sides of a previous equation by $\frac{\hbar}{im}$ and introducing the following notation  
\begin{eqnarray}\label{Z_psi}
Z(x)=\frac{\hbar}{im}\left(\frac{\psi'(x)}{\psi(x)}\right)
\end{eqnarray}
we get a first order differential equation for a quantum wave impedance
\begin{eqnarray}\label{Zequation}
\frac{dZ(x)}{dx}+i\frac{m}{\hbar}Z^2(x)=i\frac{2}{\hbar}\left(E-U(x)\right).
\end{eqnarray}
Notice that this equation is a nonlinear Riccati equation. $Z(x)$ is called a quantum wave impedance. A multiplier $\frac{\hbar}{im}$ is introduced in due to the expression for the probability current density would has the simplest form:
\begin{eqnarray}\label{j_through_Z}
j=\frac{\hbar}{2im}\left(\psi^*\frac{d\psi}{dx}-\psi\frac{d\psi^*}{dx}\right)=\frac{1}{2}|\psi(x)|^2\left(Z(x)+Z^*(x)\right)
=|\psi(x)|^2\Re \left[Z(x)\right].
\end{eqnarray}
Using (\ref{Z_psi}) we get the relation between a quantum wave impedance and a wave function in the following form
\begin{eqnarray}
\psi(x)=C\exp\left[\frac{im}{\hbar}\int Z(x) dx\right],
\end{eqnarray}
where $C$ is the constant. 
Thus, we can express the probability current density only through a quantum wave impedance function $Z(x)$:
\begin{eqnarray}\label{jZext}
j=\left|C\exp\left[\frac{im}{\hbar}\int Z(x)dx\right]\right|^2\Re\left[Z(x)\right]=|C|^2\exp\left[-\frac{2m}{\hbar}\int \Im\left[Z(x)\right]dx\right]\Re\left[Z(x)\right].
\end{eqnarray}
Now a question arise. What is the advantage of a non-linear first order equation (Ricatti type equation for a quantum wave impedance) over a linear second order equation (Shr\"{o}dinger equation)?
The main advantage is that the solution of a first-order differential equation contains only one undefined constant while a second order one has two undefined constants. So the procedure of a consecutive calculation of a quantum wave impedance is much easier compare to a wave function calculation because instead of two matching conditions it is enough to use only one.

If we define the value for a quantum wave impedance function  $Z(x)$ at an arbitrary point $x=x_0$ (we call it a boundary condition) then its values at other points are defined by the equation (\ref{Zequation}) and corresponding matching conditions. 
Generally saying we can introduce two functions: $Z^+(x)$ and $Z^-(x)$, namely
\begin{eqnarray}\label{Z+Z-}
Z^-(x,x_0),\: x<x_0,\qquad
Z^+(x,x_0), \: x>x_0,
\end{eqnarray}
so both of them belong to the solutions of the equation (\ref{Zequation}) and it is not necessarily the equality $Z^-(x_0,x_0)=Z^+(x_0,x_0)$ is fulfilled. It is also not necessarily to have the same value of $x_0$ for both $Z^-$ and $Z^+$. Often there is a need to consider the following functions
\begin{eqnarray}\label{Z+Z-n}
Z^-(x,b),\: x<b,\qquad
Z^+(x,a),\: x>a,
\end{eqnarray}
where $b>a$. The role of these two functions $Z^-(x,b)$ and $Z^-(x,a)$ we will see later.

\section{Scattering case}
Dealing with a propagation of a quantum particle we can discern two cases: a scattering case and a bound states one. In this section we will show how to use the obtained equation (\ref{Zequation}) for solving a scattering task.

We say that we have a scattering case if such $x_0$ exists that the wave incidenting on the left has the value of an energy bigger than the potential energy at any point of the area $(-\infty, x_0)$. In a case of the way incidenting on the right the condition of a scattering case is very similar: $\exists x_0, \forall x\geq x_0, U(x)<E$. Notice that these conditions do not take into account infinite periodic systems which have to be considered separately.

A solution of the equation (\ref{Zequation}) in a scattering case exists for any value of $E$ and relates the probability current density of an incidenting wave with the reflected and transmitted ones. The situation of a particular interest is the case of a resonant tunnelling when the reflected probability current density is equal to zero.

On the base of a formula (\ref{jZext}) we see that for having a constant value of a probability current density $j$ in the whole region the value of
\begin{eqnarray}
\ln\left(\Re[Z(x)]\right)-\frac{2m}{\hbar}\int\Im[Z(x)]dx
=const
\end{eqnarray}
has to be constant. This expression relates a real and an imaginary part of a quantum wave impedance. If $const=0$ we have a full reflection in a studied system or a bound state. At the same time the following relation
\begin{eqnarray}
\ln\left(\Re[Z(x)]\right)-\frac{2m}{\hbar}\int\Im[Z(x)]dx
=\ln(z_0)
\end{eqnarray} 
is a condition for finding resonant levels, where $z_0$ is a characteristic impedance of the region $x<x_0$, $x_0\rightarrow-\infty$, when a wave incidents on the left. Notice that $z_0$ is real and it is reasonable to write the previous condition in the following form
\begin{eqnarray}\label{j_const_cond}
\ln\left(\frac{\Re[Z(x)]}{z_0}\right)=\frac{2m}{\hbar}\int
\Im[Z(x)]dx.
\end{eqnarray}  
Potentials which satisfy this condition (\ref{j_const_cond}) for an arbitrary energy value (within scattering case) are called reflectionless potentials \cite{Nogami_Toyama:1998}. For an arbitrary potential we can find only  discrete energy values which satisfy this condition (\ref{j_const_cond}). But it is not convenient to calculate these energy values using (\ref{j_const_cond}).
So it would be good to find the easier way for finding resonant levels. 
For this let's consider the following potential
\begin{eqnarray}\label{U1f2}
U(x)=\left\{\begin{array}{cc}
U_1, & x\le a\\
f(x),& a<x<b \\
U_2, &  x\ge b
\end{array}\right.
\end{eqnarray}
and two quantum wave impedance functions $Z^-(x,b)$ and $Z^+(x,a)$ (remind (\ref{Z+Z-n})). Then in a case of having resonant levels for each $x_0 \in [a,b]$ we get:
\begin{eqnarray}
\ln\left(\frac{\Re[Z^+(x_0,a)]}{z_0}\right)=\frac{2m}{\hbar}\int
\Im[Z^+(x,a)]dx\Biggm|_{x=x_0},\nonumber\\
\ln\left(\frac{\Re[Z^-(x_0,b)]}{z_0}\right)=\frac{2m}{\hbar}\int
\Im[Z^-(x,b)]dx\Biggm|_{x=x_0}.
\end{eqnarray}
Subtracting one equation from another we obtain that
\begin{eqnarray}
\ln\left(\frac{\Re[Z^+(x_0,a)]}{\Re[Z^-(x_0,b)]}\right)=\frac{2m}{\hbar}\int
\Im[Z^+(x,a)-Z^-(x,b)]dx\Biggm|_{x=x_0}\!\!\!\!\!\!\!.
\end{eqnarray}
Now taking into account that this equation should be valid for each $x_0\in[a,b]$ and that in a case of a wave incidenting on the left the following equation holds $Z^+(a,a)=z_0$, we get
\begin{eqnarray}
\Re[Z^+(x_0,a)]=\Re[Z^-(x_0,b)],\!\!\!\!\quad \Im[Z^+(x_0,a)]=\Im[Z^-(x_0,b)]
\end{eqnarray}
or finally
\begin{eqnarray}\label{Res_cond}
Z^-(x_0,b)=Z^+(x_0,a), \quad x_0 \in [a,b].
\end{eqnarray}
Although we derived this condition for a potential of a form (\ref{U1f2}) it is still valid for a scattering problem with an arbitrary potential.

It is worth to say that the resonant tunnelling of waves is important in the formation of the characteristics of wave structures, especially a spectral selection which allows providing extremely high signal decoup\-ling in operating and non-operating areas of frequency and energy.

\section{Bound states case}

If such $x_1$ and $x_2$ exist that the energy of the wave is less than the potential energy in each point of the region: $x\in (-\infty, x_1)\cup (x_2,\infty)$ then we have a problem which is qualitatively different from the scattering one. It is a bound states case but note that this condition does not include infinite periodic systems 
which have to be considered separately.  The solution of the equation (\ref{Zequation}) exists for any value of energy $E$, but only for descrete values of $E$ the solution for a quantum wave impedance function $Z(x)$ corresponds to a wave function $\psi(x)$ which can be normalized per unit.  
To normalize a wave function, it is necessary at least $\psi(x)\sim x^\alpha$ with $\alpha<-1$ when  $x\rightarrow \pm\infty$. It means that 
$\frac{im}{\hbar}Z(x)x$ must be less than $-1$, when $x\rightarrow\pm\infty$:
\begin{eqnarray}\label{Z_cond_inf}
\frac{im}{\hbar}Z(x)x<-1,\quad x\rightarrow\pm\infty.
\end{eqnarray}  

The condition for finding eigenvalues of an energy is the same that in the scattering case (\ref{Res_cond}). We will demonstrate this using the potential energy of the form (\ref{U1f2}).
Assume that we have solutions of Shr\"{o}dinger equation 
\begin{eqnarray}
-\frac{\hbar^2}{2m}\frac{d^2\psi(x)}{dx^2}+U(x)\psi(x)=E\psi(x)
\end{eqnarray}
in the region $[a,b]$ both to the right and to the left of the arbitrary point $x_0 \in [a,b]$ and
$\Psi_+(x)$ is the solution to the left of the point $x_0$, 
$\Psi_-(x)$ is the solution to the right of the same point. Both solutions can be depicted as a linear combination of two linearly independent functions:
\begin{eqnarray}
\Psi_+(x_0,x)=A_+\psi(x)+B_+\phi(x),\nonumber\\
\Psi_-(x_0,x)=A_-\psi(x)+B_-\phi(x).
\end{eqnarray}
When $x\leq a$ we have
\begin{eqnarray}
\psi_L(x)=C_L\exp[\varkappa_1(x-a)],
\end{eqnarray}
and for $x\geq b$
\begin{eqnarray}
\psi_R(x)=C_R\exp[-\varkappa_2(x-b)],
\end{eqnarray}
where wave numbers $\varkappa_1=\sqrt{2m(E-U_1)}/\hbar$, $\varkappa_2=\sqrt{2m(E-U_2)}/\hbar$.

Applying boundary conditions at points $a$ and $b$ one defines
$A_+$, $B_+$, $A_-$, $B_-$ through $C_R$ and $C_L$ because one has to satisfy matching both a wave function and its first derivative at each interface. So
\begin{eqnarray}
A_+\psi(a)+B_+\phi(a)=C_L,\quad
A_+\psi'(a)+B_+\phi'(a)=\varkappa_1C_L
\end{eqnarray}
and
\begin{eqnarray}
A_-\psi(b)+B_-\phi(b)=C_R,\quad
A_-\psi'(b)+B_+\phi'(b)=\varkappa_2C_R.
\end{eqnarray}
Thus
\begin{eqnarray}
\Psi_+(x_0,a)=C_L\left(\frac{\phi'(a)-\phi(a)\varkappa_1}
{\psi(a)\phi'(a)-\phi'(x)\psi(a)}\psi(x_0)+
\frac{\psi'(a)-\psi(a)\varkappa_1}
{\psi(a)\phi'(a)-\phi'(x)\psi(a)}\phi(x_0)\right),\nonumber\\
\Psi_-(x_0,b)=C_R\!\!\left(\frac{\phi'(b)-\phi(b)\varkappa_2}
{\psi_-(b)\phi_-'(b)-\phi'(b)\psi(b)}\psi(x_0)+\frac{\psi'(b)-\psi(b)\varkappa_2}
{\psi(b)\phi'(b)-\phi'(b)\psi(b)}\phi(x_0)\right)\!\!.
\end{eqnarray}
Now at a point $x_0\in [a,b]$ one has to satisfy the following matching conditions
\begin{eqnarray}
\Psi_+(x_0,b)=\Psi_-(x_0,a),\quad
\Psi_+'(x_0,b)=\Psi_-'(x_0,a).
\end{eqnarray}
Notice that these conditions are valid only if a potential  at a point $x_0$ is not singular. In a case of a zero-range singular (at a point $x_0$) potential we would have another matching conditions which depend on the type of singularity.

The solution for $C_L$ and $C_R$ exists only if
\begin{eqnarray}
\frac{\Psi_+(x_0,b)}{\Psi_+'(x_0,b)}=
\frac{\Psi_-(x_0,a)}{\Psi_-'(x_0,a)}.
\end{eqnarray} 
It is a condition for finding eigenvalues of energy.
Multiplying both sides on $\frac{\hbar}{im}$ one gets:
\begin{eqnarray}
Z^-(a,x_0)=Z^+(b,x_0),
\end{eqnarray}
where $x_0$ is an arbitrary point from a region $(a,b)$.

It is not unexpected result since a standing wave is a superposition of two waves of equal amplitudes propagating in opposite directions. This fact physically explains the equality of impedances $Z^-$ and $Z^+$ at an arbitrary point $x_0\in (a,b)$.

\section{Solution for a constant value of potential energy}
In this section we are going to find a solution of the equation (\ref{Zequation}) in a region in which the potential energy is constant. Assume that the potential energy is equal to $U_0$ and the total energy of a particle is equal to $E$. This is one of the simplest cases but at the same time it is very important and it has many applications especially in the area of numerical calculations.

If $U(x)=U_0$ we can separate variables in the equation (\ref{Zequation}) and get that
\begin{eqnarray}
\frac{i\hbar}{m}\frac{dZ}{Z^2-z_0^2}=dx,
\end{eqnarray}
where $z_0=\sqrt{2(E-U_0)/m}$ is the characteristic impedance of the region with constant potential energy $U_0$. After integration of both parts of the previous relation  
\begin{eqnarray}
\frac{\hbar}{imz_0} \arcth\left(\frac{Z(x)}{z_0}\right)=x+x_0,
\end{eqnarray}
it is easy to get a solution for a quantum wave impedance function
\begin{eqnarray}\label{Z_const_sol}
Z(x)=z_0\th\left(\gamma_0x+\phi\right),
\end{eqnarray}
where $\gamma_0=\frac{imz_0}{\hbar}$ is the magnitude, which characterizes the wave propa\-gating in the midium with
the characteristic impedance $z_0$, $\phi$ is the constant phase which depends on the boundary conditions for a quantum wave impedance function. Generally, $\gamma_0$ is a complex quantity, and we also use both $k_0\ (ik_0=\gamma_0)$ when $\gamma_0$ is fully imaginary and $\varkappa_0\ (\varkappa_0=\gamma_0)$ in a case of $\gamma_0$ is fully real.

Let's consider the behaviour of $Z(x)$ when $x\rightarrow \pm \infty$ in two cases: $\gamma_0$ is real and $\gamma_0$ is imaginary.  
If $\gamma_0$ is real then 
\begin{eqnarray}\label{Zpminf}
\lim\limits_{x\rightarrow\pm\infty} Z(x)=\lim\limits_{x\rightarrow\pm\infty} z_0\th\left(\gamma_0 x+\phi\right)=\pm z_0=\pm\frac{\hbar}{im} \gamma_0.
\end{eqnarray}
In this case we have the attenuation of a wave function when $x\rightarrow\pm\infty$ and the condition which we obtained in a previous section is satisfied
\begin{eqnarray}
\lim_{x\rightarrow\pm\infty}\frac{im}{\hbar}Z(x)x<-1.
\end{eqnarray}
Result (\ref{Zpminf}) for the potential of the form (\ref{U1f2}) means that 
the boundary conditions for $Z(x), x\in [a,b]$ are:
\begin{eqnarray}
Z(a)=-z_1,\qquad Z(b)=z_2,
\end{eqnarray}
where $z_1=\sqrt{2(E-U_1)/m}>0$ and 
$z_2=\sqrt{2(E-U_2)/m}>0$ are the characteristic impedances of the regions $x<a$ and $x>b$ consequently.

Really, a wave function $\psi(x)=C\exp\left[\frac{im}{\hbar}\int Z(x) dx\right]$ can be normalized only if we choose $Z(a)=-z_1$ and $Z(b)=z_2$ because it gives $\psi(x)\rightarrow 0$, when $x\rightarrow\pm\infty$.    

If $\gamma_0$ is imaginary then after an appropriate representation of $\th(x)$ function it becomes obvious that the solution (\ref{Z_const_sol}) contains a linear combination of plane waves which propagate in the opposite directions:
\begin{eqnarray}
Z(x)=z_0\frac{\exp[ik_0x+\phi]-\exp[-ik_0x+\phi]}{\exp[ik_0x+\phi]-\exp[-ik_0x+\phi]},
\end{eqnarray}
where $ik_0=\gamma_0$ ($k_0$ is real).
For the potential energy of the form (\ref{U1f2}) when the wave incidents on the left there are no reasons for $Z(x)$ to be dependent on $x$, when $x\geq b$. It is possible if we put $\phi=\pm\infty$. But only $\phi=\infty$ is acceptable because only then we have the wave moving in the positive direction of $x$ axis.
\begin{eqnarray}
\psi(x)=\exp\left[\frac{im}{\hbar}\int z_0 dx\right]=C\exp[ik_2 x],\quad x\geq b,
\end{eqnarray}
where $ik_2=\gamma_2$ ($k_2$ is real).
Really, as we remember $r=\exp[-2\phi]$ in a case of a wave which incidents on the left. If $\phi=\infty$ then we have $r=0$ and $t=1$. It means that a wave is moving without reflection, which is correct for the region $x>b$.
As a result of this consideration we get the right-side boundary condition for a quantum wave impedance function: $Z(b)=z_2>0$.

The same consideration gives us that in a case of wave incidenting on the right we have to choose $\phi=-\infty$. Thus, the left-side boundary condition is $Z(a)=-z_1<0$. Indeed, in a case of $\phi=-\infty$ we have $r=\infty$ which means that there is no any wave moving in the positive direction of $x$, so a wave is moving to the left without any reflection, which is correct for the region $x<a$.

\section{Transmission and reflection coefficients}

The quantitative expression of a wave scattering are two magnitudes, namely, a wave amplitude reflection coefficient $r$ and a wave amplitude transition coefficient $t$. A wave amplitude reflection coefficient describes how much of an incidenting wave is reflected. It is equal to the ratio of an amplitude of the reflected wave to an amplitude of an incidenting wave. A wave transmission amplitude coefficient relates the amplitude of a transmitted wave with an amplitude of an incidenting wave.

Having $r$ and $t$ we can calculate both reflection $R=|r|^2$ and transmission $T=|t|^2$ probabilities. The other definition of these magnitudes is as follows:

\begin{eqnarray}
T=\frac{|j_+|}{|j|},\qquad R=\frac{|j_-|}{|j|},
\end{eqnarray}
where $j$ is a probability density current of an incidenting wave, $j_+$ is a probability density current of a  transmitted wave and $j_-$ is a probability density current of a reflected wave.

Let's consider a very simple example, namely, a potential step, and find $r$, $t$, $R$, $T$ for this case. So we have a potential energy in a following form
\begin{eqnarray}\label{Ustep}
U(x)=\left\{\begin{array}{cc}
U_1, & x\le x_0\\
U_2>U_1, &  x>x_0
\end{array}\right.
\end{eqnarray} 
and assume that the wave incidenting on the left has an energy $U_1<E<U_2$. There are two regions of a constant potential and each of them has its characteristic impedance ($z_1$ or $z_2$, $z_1$ is real and $z_2$ is imaginary) and a wave vector ($k_1$ and $\varkappa_2$, $k_1$ and $\varkappa_2$ are real). In these two regions we calculate the probability density  current on the base of a formula (\ref{j_through_Z}) and taking into account that $\phi=\phi_R+i\phi_I$ we get
\begin{eqnarray}
\!\!\!\!\!\!\!\!\!Z_1(x)\!=\!z_1\th\left(ik_1x+\phi_1\right)\!\!\!
&=&\!\!\!\frac{z_1}{2}\frac{\sh(2\phi_R)}{\sh^2(\phi_R)\!+\!\cos^2(k_1x+\phi_I)}\nonumber\\
\!\!\!&+&\!\!\!
i\frac{z_1}{2}\frac{\sin(2(k_1x+\phi_I))}{\sh^2(\phi_R)+\cos^2(k_1x+\phi_I)}
\end{eqnarray}
and
\begin{eqnarray}
\int \Im\left[Z_1(x)\right]dx=-\frac{z_1}{k_1}\ln\left[\sh^2(\phi_R)+\cos^2(k_1x+\phi_I)\right].
\end{eqnarray}
So, finally
\begin{eqnarray}\label{jj-}
|j|-|j_-|=z_1\frac{|C|^2}{2}\sh\left(2\phi_R\right)=z_1|\tilde{C}|\left(1-\exp(-4\phi_R)\right),
\end{eqnarray}
where $|\tilde{C}|^2=\frac{1}{2}|C|^2\exp(2\phi_R)$.

Value of $\phi_R$ we find from the matching condition at the $x_0$ point:
\begin{eqnarray}
z_1\th(ik_1x_0+\phi)=z_2.
\end{eqnarray}
Taking into account that
\begin{eqnarray}
\th(ik_1x_0\!+\!\phi)\!=\!\frac{1\!-\!\exp[-2(ik_1x_0+\phi)]}
{1\!+\!\exp[-2(ik_1x_0\!+\!\phi)]}\!=\!\frac{1\!-\!r\exp[-2ik_1x_0]}{1\!+\!r\exp[-2ik_1x_0]}
\end{eqnarray}
we get
\begin{eqnarray}\label{rz1z2}
r=\exp(-2\phi)=\exp[2ik_1x_0]\frac{1-z_2/z_1}{1+z_2/z_1}
\end{eqnarray}
and
\begin{eqnarray}
|r|=\exp(-2\phi_R)=\left|\frac{1-z_2/z_1}{1+z_2/z_1}\right|.
\end{eqnarray}
In the considered case $U_1<E<U_2$ the value of $|r|$ is always equal \mbox{to $1$} besides one particular case $z_1=z_2$ when $r=0$. It is not a surprising result because if $z_1=z_2$ we have a constant potential in a whole region which means that there are no reasons the wave is reflected at the {\mbox point $x_0$.}
But obtained results are valid also for the case of $U_1<U_2<E$ when both $z_1$ and $z_2$ are real. In this case $|z|$ is not equal to $1$.

On the base of obtained results we can find expressions for both a transmission probability $T$ and a reflection probability $R$, namely
\begin{eqnarray}
R=\exp(-4\phi_R),\qquad T=1-\exp(-4\phi_R)
\end{eqnarray}
or 
\begin{eqnarray}
R=\left|\frac{z_2-z_1}{z_2+z_1}\right|^2,\qquad T=1-R. 
\end{eqnarray}
Now we can find an absolute value of a wave transmission amplitude coefficient 
\begin{eqnarray}
|t|=\sqrt{1-|r|^2}.
\end{eqnarray}
But the question about a phase of $t$ is still actual.
Generally if we are interested in a  phase of $t$ at the same point a wave is reflected then we have very simple relation, namely
\begin{eqnarray}
t=1+r.
\end{eqnarray}
But if one considers the potential energy of a form (\ref{U1f2}) and
is interes\-ted in a phase of $t$ at a point $x=b$ then the previous formula is not valid (it is valid only for a point $x=a$ if we consider the wave incidenting on the left). To find the correct formula we have to use the following relations, which express the matching conditions for the wave function:
\begin{eqnarray}
C_1\exp\left[\frac{im}{\hbar}\int Z_1(x)dx\right]\biggm|_{x=a}\!\!\!&=&\!\!\!
C_2\exp\left[\frac{im}{\hbar}\int Z_2(x)dx\right]\biggm|_{x=a},\nonumber\\
C_2\exp\left[\frac{im}{\hbar}\int Z_2(x)dx\right]\biggm|_{x=b}\!\!\!&=&\!\!\!
C_3\exp[ik_3b].
\end{eqnarray}
Reminding that $t=C_3/(C_1\exp[\phi_1])$ we get
\begin{eqnarray}\label{tZ}
t\!=\!\frac{\exp\left[\frac{im}{\hbar}\int Z_1(x)dx\right]\!\!\bigm|_{x=a}\!\!\exp\left[\frac{im}{\hbar}\int Z_2(x)dx\right]\!\!\bigm|_{x=b}}
{\exp\left[\frac{im}{\hbar}\int Z_2(x)dx\right]\bigm|_{x=a}\exp[ik_3b+\phi_1]}.
\end{eqnarray}
The multiplier $\exp[\phi_1]$ in a denominator of the expression for $t$ is there because $C_1\exp[\phi_1]$ is the amplitude of an incidenting wave.

In a case of $r$ calculation for the potential energy of a form (\ref{U1f2}) we have a formula very similar to (\ref{rz1z2}), namely
\begin{eqnarray}\label{rZl}
r=\frac{z_1-Z(a)}{z_1+Z(a)},
\end{eqnarray}
when a wave incidents on the left and
\begin{eqnarray}\label{rZr}
r=\frac{z_2-Z(b)}{z_2+Z(b)}
\end{eqnarray}
when a wave incidents on the right.

\section{Tunelling through a single rectangular barrier}
A solution of a scattering task for a single  barrier plays a significan role in the development of devices of signal processing on different waves. We speak about single barrier resonant-tuneling structures which have high selective properties. In reality a tunnel barrier is realised by a semiconductor superlattice, photonic or phononic crystal \cite{Nelin_Sergiyenko:2008}. 

The simplest case of a single barrier system is a rectangular potential barrier:
\begin{eqnarray}\label{Usrb}
U(x)=\left\{\begin{array}{cc}
U_b, & a\leq x\leq b\\
0, &  (x<a) \wedge (x>b)
\end{array}\right..
\end{eqnarray}
A quntum wave impedance at the point $x=a$ can be calculated on the base of the results of a previous section. Thus,
\begin{eqnarray}
Z(a)=z_b\frac{z_0\ch(\gamma_bl)-z_b\sh(\gamma_bl)}
{z_b\ch(\gamma_bl)-z_0\sh(\gamma_bl)},
\end{eqnarray}
where $l=b-a$, $i\gamma_b=(m/\hbar)z_b$, $z_b=\sqrt{(E-U_b)/m}$, $z_0=\sqrt{E/m}$. Notice that $\gamma_b$ is real ($z_b$ is imaginary) when $E\leq U_0$ and $\gamma_b$ is imaginary ($z_b$ is real) in the opposite case.

So now on the base of formula (\ref{rZl}) we easily get the following relation: 
\begin{eqnarray}\label{r_exp}
r=\frac{(1-z_b^2/z_0^2)\sh(\gamma_bl)}
{2z_b/z_0\ch(\gamma_bl)-(1+z_b^2/z_0^2)\sh(\gamma_bl)}=\frac{(1-\gamma_b^2/\gamma_0^2)\sh(\gamma_bl)}
{2\gamma_b/\gamma_0\ch(\gamma_bl)-(1+\gamma_b^2/z_0^2)\sh(\gamma_bl)}.
\end{eqnarray}

To find $R$ it is worth to consider two cases separately: $E\leq U_b$ and
$E>U_b$. In the first case ($E\leq U_b$) we have that $\gamma_b$ and $z_0$ are real, $z_b$ is imaginary. So after introducing $k_0=\frac{m}{\hbar}z_0$ ($ik_0=\gamma_0$, $\gamma_0$ is imaginary) 
and $\varkappa_b=\frac{mz_b}{i\hbar}=\sqrt{2m(U_b-E)}/\hbar>0$($z_b$ is imaginary) 
we get
\begin{eqnarray}
R=|r|^2=
\frac{(1+\varkappa_b^2/k_0^2)^2\sh^2(\varkappa_bl)}
{4\gamma_b^2/k_0^2+(1+\varkappa_b^2/k_0^2)^2\sh^2(\varkappa_bl)}.
\end{eqnarray}
In the second case ($E>U_0$) we have that $\gamma_b$ is imaginary, $z_b$ and $z_0$ are real. So after introducing $k_b=\frac{m}{\hbar}z_b$ ($ik_b=\gamma_b$) we get
\begin{eqnarray}
R=|r|^2=
\frac{(1-k_b^2/k_0^2)^2\sin^2(k_bl)}
{4k_b^2/k_0^2+(1-k_b^2/k_0^2)^2
	\sin^2(k_bl)}.
\end{eqnarray}
To find $t$ we can use formula (\ref{tZ}), in which for the simplification of calculations we assume $a=0$: 
\begin{eqnarray}\label{trectb}
t\!=\!\frac{\exp\!\!\left[\gamma_0\!\int\! \th(\gamma_0x\!+\!\phi_1)dx\right]\!\!\bigm|_{x=0}\!\exp\!\!\left[\gamma_b\!\int\! \th(\gamma_bx\!+\!\phi_2)dx\right]\!\!\bigm|_{x=b}}
{\exp\left[\gamma_b\int \th(\gamma_bx+\phi_2)dx\right]\bigm|_{x=0}\exp[\gamma_0b+\phi_1]}
.\end{eqnarray}
So we get
\begin{eqnarray}
t=\frac{\ch(\phi_1)\ch(\gamma_bb\!+\!\phi_2)}
{\ch(\gamma_bb\!+\!\phi_2)e^{\gamma_0b+\phi_1}}=\frac{1}{2}(1+e^{-2\phi_1})e^{-\gamma_0b}\left(\ch[\gamma_bb]+\sh[\gamma_bb]\th[\phi_2]\right).
\end{eqnarray}
Values of $e^{-2\phi_1}$ and $\th[\phi_2]$ we get on the base of matching conditions for a quantum wave impedance:
\begin{eqnarray}
-z_0\th(\phi_1)=z_b\th(\phi_2);\quad
z_b\th(\gamma_bb+\phi_2)=z_0.
\end{eqnarray}
Reminding that the second matching condition can be written as follows 
\begin{eqnarray} 
z_b\frac{\th(\gamma_bb)+\th(\phi_2)}{1+\th(\gamma_bb)\th(\phi_2)}=z_0
\end{eqnarray}
we find that
\begin{eqnarray}
&&\th{\phi_2}=\frac{z_0/z_b-\th(\gamma_bb)}{1-z_0/z_b\th(\gamma_bb)},
\nonumber\\
&&1+\exp[-2\phi_1]=\frac{2z_0}{z_1\th[\phi_2]+z_0}=\frac{(1-z_0/z_b\th(\gamma_bb))}{1-(z_b^2+z_0^2)/(2z_0z_b)\th(\gamma_bb)}.
\end{eqnarray}
Substituting it into the initial relation and taking into account that $b=l$ (because of $a=0$ assumption) we finally get
\begin{eqnarray}
t=\frac{2z_0z_b\exp[-\gamma_0l]}{2z_0z_b\ch(\gamma_bl)-(z_0^2+z_b^2)\sh(\gamma_bl)}=\frac{2\gamma_0\gamma_b\exp[-\gamma_0l]}{2\gamma_0\gamma_b\ch(\gamma_bl)-(\gamma_0^2+\gamma_b^2)\sh(\gamma_bl)}.
\end{eqnarray}
Notice that $\gamma_0$ is imaginary regardless on the value of $U_b$.

When $E<U_b$ we have that $z_0$ and $\varkappa_b$ are real and $z_b$ is imaginary. Thus, we have
\begin{eqnarray}
T=|t|^2=\frac{4k_0^2\varkappa_b^2}
{4k_0^2\varkappa_b^2\ch^2(\varkappa_bl)+(k_0^2-\varkappa_b^2)^2\sh^2(\varkappa_bl)}
\left\{1+\frac{1}{4}\left(\frac{k_0}{\varkappa_b}+\frac{\varkappa_b}{k_0}\right)^2\sh^2(\varkappa_bl)\right\}^{-1}\!\!\!\!\!\!.
\end{eqnarray}
For $E>U_0$,  $z_0$ and $z_b$ are real and $\gamma_b$ is imaginary, $ik_b=\gamma_b$ and
\begin{eqnarray}
T=|t|^2=\frac{4k_0^2k_b^2}
{4k_0^2k_b^2\cos^2(k_bl)\!+\!(k_0^2+k_b^2)^2\sin^2(k_bl)}=
\left\{1\!+\!\frac{1}{4}\left(\frac{k_0}{k_b}\!-\!\frac{k_b}{k_0}\right)^2\!\!\sin^2(k_bl)\right\}^{-1}\!\!\!\!\!\!.
\end{eqnarray}

\section{Conclusions}
Very often the most important and interest aspect (especially in the area of a practical application) of a wave propagation is the issue of an energy transmission. This includes a calculation of such parameters as transmission and reflection coefficients and finding conditions of a formation of standing waves and resonant states. In this case an impe\-dance concept is very fruitful because it gives the possibility to calculate the mentioned parameters in an easier way than by other methods.

This is clearly seen in the example of the calculation of a reflection coefficient for a singular rectangular barrier. We got the result for $r$ (\ref{r_exp}) very easy by using a quantum wave impedance approach.
Other approaches, namely classical method based on a direct solving of a Shr\"{o}dinger equation and a transfer matrix technique demand much more efforts to complit this task. In this case a classical method implies constructing and solving 2 determinants of $4\times 4$ size while within a transfer matrix method we have to build three matrixes of $2\times 2$ and then myptiply them in a correct order. 

What is the reason of this advantage? As we mentioned earlier the main advantage is that the solution of a first-order differential equation (for a quantum wave impedance function) contains only one undefined constant while a second order equation (Shr\"{o}dinger) has two undefined constants. Thus, a consecutive calculation of a quantum wave impedance is much easier compare to a wave function calculation because instead of two matching conditions it is enough to use only one.

All this means that a quantum wave impedance approach is very effective tool for a calculation of
probability amplitudes of a particle reflection and transition, especially in a case of systems with many wells and barriers, finding the conditions of a resonant tunneling and bound states. 

Concluding, we have to say that in general the application of impedance models for quantum-mechanical systems allows simplifying and generalizing (in comparison with other approaches) the process of these systems studying significantly.
This clearly can be seen from the articles \cite{Nelin:2007, Bojko_Berezjanskyi_Nelin:2007, Nelin_Vodolazka:2014, Akhmedov_Nelin:2007}.

\renewcommand\baselinestretch{1.0}\selectfont


\def\name{\vspace*{-0cm}\LARGE 
	Bibliography\thispagestyle{empty}}
\addcontentsline{toc}{chapter}{Bibliography}

{\small

	\bibliographystyle{gost780u}
	\bibliography{full.bib}
	
}

\newpage

\end{document}